\documentclass[11pt]{article}
\usepackage{authblk}
\usepackage{amsfonts}
\usepackage{amsmath}
\usepackage{amsthm}
\usepackage{amssymb}
\usepackage{amscd}
\usepackage{eucal}
\usepackage{bm}
\usepackage{graphicx}
\usepackage{float}

\usepackage{tikz}
\usepackage{tikz-3dplot}
\usetikzlibrary{shadows}
\usetikzlibrary{positioning}
\usetikzlibrary{arrows.meta}
\usepackage[colorlinks = true,
            linkcolor = blue,
            urlcolor  = blue  ,
            citecolor = blue,
            anchorcolor = blue]{hyperref}
\usepackage{braket}
\usepackage{xcolor}
\usepackage{subcaption}
\usepackage{pgfplots}
\usepackage{times}

\usepackage[a4paper, total={7in, 9.5in}, top=1.2in]{geometry}
\usepackage[style=numeric-comp,maxnames=9,sorting=none,backend=biber]{biblatex}
\AtEveryBibitem{%
  \clearfield{note}%
  \clearfield{reportNumber}%
}

\usepackage[utf8]{inputenc}

\usepackage{graphicx}
\usepackage[colorlinks = true,
            linkcolor = blue,
            urlcolor  = blue  ,
            citecolor = blue,
            anchorcolor = blue]{hyperref}
\usepackage{braket}
\usepackage{xcolor}

\newcommand{\eE}{\mathrm{e}}
\newcommand{\el}{\ell}
\def\i{{\rm i}}

\begin{document}

\title{Non-linear integral equations for the XXX spin-1/2 quantum chain with
non-diagonal boundary fields}

\bigskip
\author[1]{Holger Frahm}%
\affil[1]{Institut für Theoretische Physik, Leibniz Universität Hannover, Appelstr. 2, 30167 Hannover,Germany}
\author[2]{Andreas Kl\"umper} 
\affil[2]{Department of Physics,
 University of Wuppertal, Gaussstra\ss e 20, 42119 Wuppertal,
  Germany}
\author[2]{Dennis Wagner}
\author[3]{Xin  Zhang}
\affil[3]{Beijing National Laboratory for Condensed Matter Physics, Institute of Physics, Chinese Academy of Sciences, Beijing 100190, China}

\date{\today}
\maketitle

\begin{abstract}
  The XXX spin-$\frac{1}{2}$ Heisenberg chain with non-diagonal boundary
  fields represents a cornerstone model in the study of integrable systems
  with open boundaries. Despite its significance, solving this model exactly
  has remained a formidable challenge due to the breaking of $U(1)$
  symmetry. Building on the off-diagonal Bethe Ansatz (ODBA), we derive a set
  of nonlinear integral equations (NLIEs) that encapsulate the exact spectrum
  of the model.

  For $U(1)$ symmetric spin-$\frac{1}{2}$ chains such NLIEs involve two
  functions $a(x)$ and $\bar{a}(x)$ coupled by an integration kernel with
  short-ranged elements. The solution functions show characteristic features
  for arguments at some length scale which grows logarithmically with system
  size $N$.

  For the non $U(1)$ symmetric case, the equations involve a novel third
  function $c(x)$, which captures the inhomogeneous contributions of the
  $T$-$Q$ relation. The kernel elements coupling this function to the
  standard ones are long-ranged and lead for the ground-state to a winding
  phenomenon. In $\log(1+a(x))$ and $\log(1+\bar a(x))$ we observe a
  sudden change by $2\pi$i at a characteristic scale $x_1$ of the
  argument. Other features appear at a value $x_0$ which is of order $\log N$.
  These two length scales, $x_1$ and $x_0$, are independent: their ratio
  $x_1/x_0$ is large for small $N$ and small for large $N$. Explicit
  solutions to the NLIEs are obtained numerically for these limiting cases,
  though intermediate cases ($x_1/x_0 \sim 1$) present computational
  challenges.

  This work lays the foundation for studying finite-size corrections and
  conformal properties of other integrable spin chains with non-diagonal
  boundaries, opening new avenues for exploring boundary effects in quantum
  integrable systems.
  
\end{abstract}


\section{\label{sec:level1}Introduction}

The XXX spin-\(\frac{1}{2}\) chain with periodic boundary
  conditions is a seminal representative of quantum integrable systems.  A
  mathematically sufficient condition for integrability is the Yang-Baxter
  equation~\cite{Baxter1982}, a fundamental tool that has shaped much of our
  modern understanding and has led to the dicovery of many new integrable
  systems.  While the XXX chain with periodic boundary conditions was solved
  by Bethe in the famous paper~\cite{Bethe31}, integrability in the presence
  of boundaries introduces rich and intricate challenges.

Non periodic boundary conditions significantly influence
  integrable systems, and their exploration has a long history. For instance,
  the single-component Bose gas with delta-function interactions and open
  boundaries was solved in \cite{Gaudin1971}. The XXX spin-\(\frac{1}{2}\)
  chain with parallel boundary fields has been successfully solved using both
  coordinate and algebraic Bethe Ansatz
  methods~\cite{Alcaraz1987,Sklyanin1988}, with finite-size corrections
  studied in \cite{Asakawa1996FiniteSize}.

Extending these goals to more general boundary conditions and
  models required new theoretical tools, such as Sklyanin's reflection
  algebra~\cite{Cherednik1984,Sklyanin1988}, which elegantly accounts for the
  factorization of scattering processes at the chain ends. This framework is
  the basis of the proof of integrability for the Heisenberg spin chain with
  general non-parallel boundary fields in \cite{DeVega1994}.

However, non-parallel boundary fields break the \(U(1)\) symmetry of the
system, presenting a formidable challenge to traditional Bethe Ansatz
methods. To address this, several advanced techniques have been developed.
For example, \(T-Q\) relations have been applied to specific
  cases of the partially anisotropic Heisenberg model, the XXZ chain, such as
  for root of unity cases of the anisotropy~\cite{Nepo02} or for special choices of the boundary parameters~\cite{Nepo03,YaNZ06}. Fusion techniques have also provided insights
through hierarchies of transfer matrices satisfying \(T\)- and
\(Y\)-systems~\cite{fra11}. More recently, the off-diagonal Bethe Ansatz
(ODBA)~\cite{Cao2013, Wang2016} introduced an elegant framework leveraging
commuting transfer matrices and inhomogeneous \(T-Q\)-relations. Two such
formulations of the ODBA for the XXX spin-\(\frac{1}{2}\) chain have been
developed~\cite{Wang2016}, and the completeness of one of these
approaches~\cite{Cao2013} has been argued in~\cite{Nepomechie2013}.

Alternative methods have also emerged, including the modified algebraic Bethe
Ansatz, which incorporates 
chiral
basis states into the algebraic Bethe
Ansatz
framework~\cite{CaoX03,Belliard2015,Belliard2015b,Crampe2017,Avan2015ModifiedXXZ},
the chiral coordinate Bethe ansatz~\cite{ZhangKP21}, and separation of
variables techniques, which have been applied to increasingly general boundary
conditions from~\cite{FrSW08,Niccoli2012NonDiagonalXXZ} to
~\cite{Faldella2014CompleteSpectrum}.

Recent studies have focused on deriving physical properties that are independent of the boundary field angles, such as the ground-state energy, surface energy, and low-lying excitations in the thermodynamic limit~\cite{Dong2023}. In this paper, we present a new analysis to the isotropic Heisenberg chain with arbitrary boundary fields, explicitly retaining terms that depend on the angle between the boundary fields.
Starting from the
inhomogeneous \(T-Q\)-relation with two $Q$-functions~\cite{Cao2013,Wang2016},
we derive the ground-state energy, surface energy, and finite-size
corrections, offering a comprehensive perspective on these fundamental
quantities.

The paper is organized as follows. In Section~\ref{sec:level2} we summarize
the results of the ODBA approach~\cite{Cao2013,Wang2016} which we use as our
starting point. In this paper we focus on the case of negative $p, q$
parameters (negative longitudinal components of the boundary
  fields) and leave other combinations for a future publication. In
Section~\ref{sec:level3} we identify useful auxiliary functions that satisfy a
set of functional equations. These equations are rewritten as linear equations
for the Fourier transforms of the auxiliary functions. By use of the
analyticity of the eigenvalue function these linear equations close.
Section~\ref{sec:level4} presents the analytic results for the bulk and
surface energies and numerical results for the auxiliary functions. The paper
closes with conclusions in Section~\ref{sec:level5}.

\section{\label{sec:level2}The inhomogeneous $T-Q$ relation}
For the spin-1/2 Heisenberg model with isotropic bulk interaction 
the Hamiltonian of the system with arbitrary boundary fields can be brought to
the form
\[
H =
\sum_{j=1}^{N-1}{\vec\sigma}_j\cdot{\vec\sigma}_{j+1}+\frac1p\sigma^z_1+\frac1q
\left(\sigma^z_N+\xi\sigma^x_N\right),
\]
where \(N\) is the number of sites, and \(p\), \(q\), and \(\xi\) are boundary parameters.

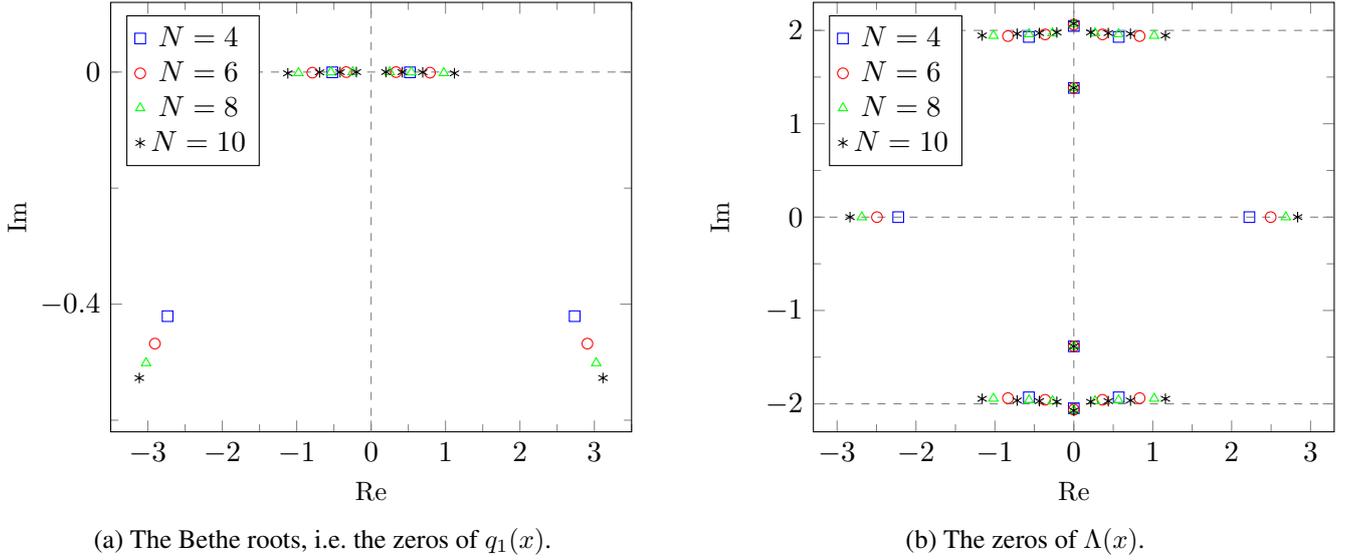
\begin{figure}
\begin{subfigure}{0.48\textwidth}
\begin{tikzpicture}
\begin{axis}[
xlabel={${\rm Re}$},
xlabel style={
font=\small},
ylabel={${\rm Im}$},
ylabel style={
font=\small},
legend pos=north west,
xmin=-3.5, xmax=3.5,
ymin=-0.62, ymax=0.12,
ytick distance=0.4,
minor tick num=1
]

\addplot[color=blue, only marks, mark=square]
  coordinates {
(2.735335,-0.420694) (-2.735335,-0.420694) ( 0.520778,-0.000481) (-0.520778,-0.000481)};
  \addlegendentry{$N=4$}

\addplot[color=red, only marks, mark=o]
  coordinates {
(2.905959,-0.467979) (-2.905959,-0.467979) (-0.789874,-0.001084) (0.789874,-0.001084) (0.334673,-0.000121) (-0.334673,-0.000122)};
\addlegendentry{$N=6$}

\addplot[color=green, only marks, mark=triangle]
  coordinates {
(3.024703, -0.501435) (-3.024703, -0.501435) (0.976258, -0.001666) (-0.976258, -0.001666) (0.541067, -0.000299) (-0.541067, -0.000299) (0.248756, -0.000049) (-0.248756, -0.000049)
};
\addlegendentry{$N=8$}

\addplot[color=black, only marks, mark=asterisk]
  coordinates {
(3.115500, -0.527200) (-3.115500, -0.527200) (1.119600, -0.002217) (-1.119600, -0.002217) (0.693180, -0.000480) (-0.693180, -0.000480) (0.417950, -0.000130) (-0.417950, -0.000130) (0.198460, -0.000025) (-0.198460,-0.000025)};
\addlegendentry{$N=10$}

\addplot[color=gray,-,dashed]
  coordinates {(0,-0.62) (0,0.42)};
\addplot[color=gray,-,dashed]
  coordinates {(-3.5,0) (3.5,0)};

\end{axis}
\end{tikzpicture}
\caption{The Bethe roots, i.e.~the zeros of $q_1(x)$.}
\label{fig:subfig1}
\end{subfigure}
\hfill 
\begin{subfigure}{0.48\textwidth}
\begin{tikzpicture}
\begin{axis}[
xlabel={${\rm Re}$},
xlabel style={font=\small},
ylabel={${\rm Im}$},
ylabel style={font=\small},
legend pos=north west,
xmin=-3.3, xmax=3.3,
ymin=-2.3, ymax=2.3,
ytick distance=1,
minor tick num=1
]

\addplot[color=blue, only marks, mark=square]
  coordinates {
(0.000000, 1.384136) (0.000000, -1.384136) (0.569216, 1.930301) (0.569216, -1.930301) (-0.569216, 1.930301) (-0.569216, -1.930301) (0.000000, 2.047488) (0.000000, -2.047488) (2.224509, 0.000000) (-2.224509, 0.000000)
};
  \addlegendentry{$N=4$}

\addplot[color=red, only marks, mark=o]
  coordinates {
(0.000000, 1.384111) (0.000000, -1.384111) (2.494285, 0.000000) (-2.494285, 0.000000) (0.363699, 1.957143) (0.363699, -1.957143) (-0.363699, 1.957143) (-0.363699, -1.957143) (0.834293, 1.938980) (0.834293, -1.938980) (-0.834293, 1.938980) (-0.834293, -1.938980) (0.000000, 2.062508) (0.000000, -2.062508)
};
\addlegendentry{$N=6$}

\addplot[color=green, only marks, mark=triangle]
  coordinates {
(0.000000, 2.069010) (0.000000, -2.069010) (0.000000, 1.384111) (0.000000, -1.384111) (2.686778, 0.000000) (-2.686778, 0.000000) (0.567448, 1.961009) (0.567448, -1.961009) (-0.567448, 1.961009) (-0.567448, -1.961009) (1.018830, 1.943366) (1.018830, -1.943366) (-1.018830, 1.943366) (-1.018830, -1.943366) (0.270093, 1.970310) (0.270093, -1.970310) (-0.270093, 1.970310) (-0.270093, -1.970310)
};
\addlegendentry{$N=8$}

\addplot[color=black, only marks, mark=asterisk]
  coordinates {
(1.161225, 1.945963) (1.161225, -1.945963) (-1.161225, 1.945963) (-1.161225, -1.945963) (0.718161, 1.964103) (0.718161, -1.964103) (-0.718161, 1.964103) (-0.718161, -1.964103) (0.437207, 1.971131) (0.437207, -1.971131) (-0.437207, 1.971131) (-0.437207, -1.971131) (0.215540, 1.978208) (0.215540, -1.978208) (-0.215540, 1.978208) (-0.215540, -1.978208) (0.000000, 2.072184) (0.000000, -2.072184) (0.000000, 1.384111) (0.000000, -1.384111) (2.836277, 0.000000) (-2.836277, 0.000000)
};
\addlegendentry{$N=10$}
\addplot[color=gray,-, dashed]
  coordinates {(-3.3,2) (3.3,2)};
\addplot[color=gray,-, dashed]
  coordinates {(-3.3,-2) (3.3,-2)};
\addplot[color=gray,-,dashed]
  coordinates {(0,-2.3) (0,2.3)};
\addplot[color=gray,-,dashed]
  coordinates {(-3.3,0) (3.3,0)};
\end{axis}
\end{tikzpicture}
\caption{The zeros of $\Lambda(x)$.}
\label{fig:subfig2}
\end{subfigure}
\caption{Depiction of Bethe roots and zeros of the eigenvalue function for the
ground-state and parameters $\{p, q,\xi\}=\{-0.6, -0.3, 1.2\}$.} 
\label{fig:zeros}
\end{figure}

We start by using the inhomogeneous $T-Q$ relation of \cite{Cao2013,Wang2016}
with two $Q$-functions. We introduce new combinations of the boundary
parameters that are useful for our purposes
\begin{equation}
\xi_1:=(1+\xi^2)^{1/2},\qquad p_1:=-2p,\qquad p_2:=-2q/\xi_1.
\end{equation}
We study the eigenvalue function for the associated transfer matrix for an
even number of sites $N$ and rescale the function by dropping a constant factor $(\i/2)^{2N+2}$
\begin{equation}
\Lambda(x)=\lambda_1(x)+\lambda_2(x)+\lambda_3(x),\label{Lambdasum}
\end{equation}
where we also reparameterized the argument $u=\i x/2-1/2$ used in \cite{Wang2016}
by the new variable $x$. The summands in (\ref{Lambdasum}) are
\begin{equation}
\lambda_1(x):=\phi_1(x)\frac{q_1(x+2\i)}{q_2(x)},\quad
\lambda_2(x):=\frac{\phi_2(x)}{q_1(x)q_2(x)},\quad
\lambda_3(x):=\phi_3(x)\frac{q_2(x-2\i)}{q_1(x)},
\end{equation}
where $q_1(x)$ and $q_2(x)$ are polynomials of degree $N$ with
$q_2(x)=q_1(-x)$. The zeros of $q_1(x)$ are called
Bethe roots. See Fig.\ref{fig:subfig1} for the depiction of zeros of $q_1(x)$ for short
system size.

The functions $\phi_1(x)$, $\phi_2(x)$, $\phi_3(x)$ are
explicitly given by
\begin{align}
\phi_1(x)&:=\xi_1\frac{\varphi(x)}x (x-\i)^{2N+1},\qquad
\varphi(x):=(x+\i+\i p_1)(x+\i+\i p_2),\nonumber\\
\phi_2(x)&:=2\big(1-\xi_1\big)(x^2+1)^{2N+1},\nonumber\\
\phi_3(x)&:=\xi_1\frac{\overline\varphi(x)}x (x+\i)^{2N+1},\qquad
\overline\varphi(x):=(x-\i-\i p_1)(x-\i-\i p_2).
\end{align}
In \cite{Wang2016} the Bethe roots are calculated for relatively short chains by
numerically solving for the Bethe equations. These equations are derived from the
condition that all potential poles in (\ref{Lambdasum}) cancel resulting in an
analytic function $\Lambda(x)$. We do not write down these equations
explicitly as we do not use them.

\section{\label{sec:level3}Transformation of functional equations to integral equations}
Here we adopt a different resp.~opposite approach to deriving Bethe ansatz
equations from the analyticity of the eigenvalue function. We avoid the
calculation of the Bethe roots and use the analyticity of $\Lambda(x)$
itself. The entire reasoning is based on identifying a set of suitable
analytic functions satisfying sufficiently many functional equations. These
will be rewritten by use of the Fourier transform into (non-linear) integral
equations of convolution type. In analogy to \cite{JK97,JKS97a}, we find that
the following functions are useful
\begin{align}
b(x)&:=\frac{\lambda_2(x)+\lambda_3(x)}{\lambda_1(x)},
&&B(x):=1+b(x)=\frac{\Lambda(x)}{\lambda_1(x)},\\
\overline b(x)&:=\frac{\lambda_1(x)+\lambda_2(x)}{\lambda_3(x)},
&&\overline B(x):=1+\overline b(x)=\frac{\Lambda(x)}{\lambda_3(x)},\\
c(x)&:=\frac{\lambda_2(x)\Lambda(x)}{\lambda_1(x)\lambda_3(x)},
&&C(x):=1+c(x)=\frac{[\lambda_1(x)+\lambda_2(x)][\lambda_2(x)+\lambda_3(x)]}{\lambda_1(x)\lambda_3(x)}.
\end{align}
Note that $b$, $\overline b$, $c$ are defined in terms of the (not explicitly
known) functions appearing in (\ref{Lambdasum}). These functions plus the
constant $1$ are called $B$, $\overline B$, $C$ and also factorize into the
same factors $\lambda_i$, $\lambda_1+\lambda_2$, $\lambda_2+\lambda_3$ as well as
 $\Lambda$.

The asymptotic behaviour of the functions for large arguments is
\begin{align}
\Lambda(x)&\simeq 2\cdot x^{2N+2},\nonumber\\
b(\infty)&=\overline b(\infty)=2/\xi_1-1,\qquad c(\infty)=4(1/\xi_1^2-1/\xi_1).\label{limits}
\end{align}
The function $\Lambda(x)$ is analytic, even and possesses a number $2N+2$ of
zeros. Numerical work for the largest eigenvalue shows that $N$ of its zeros
have imaginary parts larger than $+1$ with most of them close to $+2$, the
same number of zeros have imaginary parts less than $-1$ and most of them
close to $-2$, and two zeros $\pm x_0$ lie on the real axis. See
Fig.~\ref{fig:subfig2} for the depiction of zeros of $\Lambda(x)$ for short
system size.

By use of the above definitions and of the cancellation of poles in
(\ref{Lambdasum}) resp.~the Bethe ansatz equations we conclude that
\begin{equation}
D(x):=\frac{x q_2(x)[\lambda_2(x)+\lambda_3(x)]}{(x+\i)^{2N+1}},\quad
\overline D(x):=\frac{x q_1(x)[\lambda_1(x)+\lambda_2(x)]}{(x-\i)^{2N+1}},
\end{equation}
are polynomials of degree $N+2$.

The rational functions $b$, $\overline b$, $c$, $B$, $\overline B$, $C$ allow for the
following factorizations in terms of polynomials
\begin{align}
b(x)&=\text{cst.}\frac{(x+\i)^{2N+1}}{\varphi(x)(x-\i)^{2N+1}}\frac{D(x)}{q_1(x+2\i)},\nonumber\\
\overline b(x)&=\text{cst.}\frac{(x-\i)^{2N+1}}{\overline\varphi(x)(x+\i)^{2N+1}}\frac{\overline
  D(x)}{q_2(x-2\i)},\nonumber\\
c(x)&=\text{cst.}\frac{x^2}{\varphi(x)\overline\varphi(x)}\frac{\Lambda(x)}{q_1(x+2\i)q_2(x-2\i)},\nonumber\\
B(x)&=\text{cst.}\frac{x}{\varphi(x)(x-\i)^{2N+1}}\frac{q_2(x)}{q_1(x+2\i)}\Lambda(x),\nonumber\\
\overline
B(x)&=\text{cst.}\frac{x}{\overline\varphi(x)(x+\i)^{2N+1}}\frac{q_1(x)}{q_2(x-2\i)}\Lambda(x),\nonumber\\
C(x)&=\text{cst.}\frac{1}{\varphi(x)\overline\varphi(x)}\frac{D(x)\overline D(x)}{q_1(x+2\i)q_2(x-2\i)},\label{multiplfunceqs}
\end{align}
where in principle we may identify the (non-zero) constants, but we will not
need those.

We want to ``solve'' these functional equations of multiplicative type with
constant shifts in the arguments. The first step is the application of the
logarithm, turning the product form into additive form. In the second step
we apply the Fourier transform which turns functions with shifts in the
arguments to transforms with simple factors leading eventually to a set of
linear equations. Care has to be taken that the Fourier transform exists and
the region of convergence of the inverse transform, the Fourier
representation, is wide enough for our purposes. For these reasons we deal
with the logarithmic derivatives of the equations above upon which the
polynomial factors turn into Fourier transformable functions.

In case of the largest eigenvalue, $\Lambda(x)$ has two zeros in the strip
$|$Im$(x)|<1+\epsilon$ (where $\epsilon$ is small and positive) lying
symmetrically at points $\pm x_0$ on the real axis. The function $c(x)$ has
four zeros on the real axis, namely $\pm x_0$ and the zero at $0$ of second
order.  Hence the logarithmic derivatives of $\Lambda(x)$ and $c(x)$ do not
have Fourier representations in the strip $|$Im$(x)|<1+\epsilon$.  We therefore
introduce the analytic and non-zero functions $\widetilde\Lambda(x)$,
$\widetilde c(x)$ by
\begin{align}
\Lambda(x)&=\lambda_0(x)\widetilde\Lambda(x),&&\lambda_0(x):=(x-x_0)(x+x_0),\nonumber\\
c(x)&=x^2\lambda_0(x)\widetilde c(x).&&
\end{align}
The logarithmic derivatives of $\widetilde\Lambda(x)$ and $\widetilde c(x)$
have  Fourier representations in suffiently wide strips around the real axis.

We use the Fourier transform pair
\begin{equation}
\widetilde f(k)=\frac 1{2\pi}\int_{-\infty}^{\infty}dx\, \eE^{-\i kx}
f(x), \qquad f(x)=\int_{-\infty}^{\infty}dk\, \eE^{\i kx} \widetilde f(k),
\end{equation}
with the alternative notation $FT[f]\equiv\widetilde f$ where the standard argument
of $FT[f]$ is an implicit $k$. Note that
\begin{equation}
  f(x) \leftrightarrow \widetilde f(k) \quad \hbox{implies}\quad  f(x+\i c)
  \leftrightarrow \eE^{-ck} \widetilde f(k).
\end{equation}
We will often use the simple explicit Fourier transforms of (the logarithmic
derivatives of) linear factors
\begin{align}
FT\left[\frac{d}{dx}\log(x+\i c)\right]&=_{{\rm Im}(c)>0}
\begin{cases}
-\i\eE^{-ck},& k>0,\\
0,&k<0,
\end{cases}\label{explicitFT1}\\
&=_{{\rm Im} (c)<0}
\begin{cases}
0, &k>0,\\
\i \eE^{-ck}, &k<0.
\end{cases}
\label{explicitFT2}
\end{align}

Next we introduce a shorthand for the Fourier transform of the
logarithmic derivative of a function $f(x)$, which is very useful but
potentially misleading
\begin{equation}
f:=FT\left[\frac{d}{dx}\log f(x)\right],\label{logFTnotation}
\end{equation}
i.e.~the same symbol $f$ without specifying the argument $k$ shall mean a new
function (in general) different from $f(x)$. At first sight this looks
confusing if not non-sensic. However in practical calculations, 
misunderstandings are almost excluded: the same symbols appearing above in
multiplicative relations now appearing in additive relations have a different
meaning. This notation keeps the symbols manageable as the use of several
levels of tildes,  bars or indices is avoided.

From the above (\ref{explicitFT1},\ref{explicitFT2}) we immediately calculate the transforms for the logarithmic derivatives of
$\varphi(x)$ resp.~$\overline\varphi(x)$ (noting that $p, q <0$, i.e.~$p_1,
p_2 >0$) and by use of the notation (\ref{logFTnotation})
\begin{equation}
\varphi=
\begin{cases}
-\i (\eE^{-(1+p_1)k}+\eE^{-(1+p_2)k}),&k>0,\\
0,&k<0,
\end{cases}\qquad
\overline\varphi=
\begin{cases}
0, &k>0,\\
\i (\eE^{(1+p_1)k}+\eE^{(1+p_2)k}), &k<0,
\end{cases}\label{FTphis}
\end{equation}
The functions $q_1(x)$, $q_2(x)$, have zeros close to the real axis.  For the
logarithmic derivatives of $q_1(x)$ resp.~$q_2(x)$ we use Fourier
representations in the upper resp.~lower half plane with vanishing Fourier
components for $k<0$ resp.~$k>0$. For $p_1, p_2 >0$ and small system size the
functions $D(x)$ resp.~$\overline D(x)$ have zeros in the upper resp.~the
lower half-plane. This is also the case for all system sizes if $p_1, p_2 >1$
and certain cases of $p_1>1>p_2 >0$ which is what we consider in this paper.
In any case, for large system size the bulk of the zeros of $D(x)$
resp.~$\overline D(x)$ is close to Im$(x) =+2$ resp.~Im$(x) =-2$.  For the
logarithmic derivatives of $D(x)$ resp.~$\overline D(x)$ we use Fourier
representations in the semi-planes Im$(x) <+2$ resp.~Im$(x) >-2$ with
vanishing Fourier components for $k>0$ resp.~$k<0$.

Now we apply the described transform to (\ref{multiplfunceqs}) and obtain
\begin{align}
\hbox{for}\ k>0:\qquad b&=-(2N+1)\i\eE^{-k}-\varphi-\eE^{-2k}q_1,\nonumber\\
\overline b&=(2N+1)\i\eE^{-k}+\overline D,\nonumber\\
\widetilde c&= -\varphi-\eE^{-2k}q_1+\widetilde\Lambda,\nonumber\\
B&=-\varphi-\eE^{-2k}q_1+\widetilde\Lambda,\nonumber\\
\overline B&=-\i+(2N+1)\i\eE^{-k}-2\i\cos(k x_0)+q_1+\widetilde\Lambda,\nonumber\\
C&=-\varphi+\overline D-\eE^{-2k}q_1,\label{eqkpos}
\end{align}
and
\begin{align}
\hbox{for}\ k<0:\qquad b&=-(2N+1)\i\eE^{k}+D,\nonumber\\
\overline b&=(2N+1)\i\eE^{k}-\overline\varphi-\eE^{2k}q_2,\nonumber\\
\widetilde c&= -\overline\varphi-\eE^{2k}q_2+\widetilde\Lambda,\nonumber\\
B&=\i-(2N+1)\i\eE^{k}+2\i\cos(k x_0)+q_2+\widetilde\Lambda,\nonumber\\
\overline B&=-\overline\varphi-\eE^{2k}q_2+\widetilde\Lambda,\nonumber\\
C&=-\overline\varphi+D-\eE^{2k}q_2.\label{eqkneg}
\end{align}
The last three equations of (\ref{eqkpos}) can be solved for $q_1$, $\overline
D$, $\widetilde\Lambda$ in terms of $B$, $\overline B$, $C$ and this inserted
into the first three equations gives
\begin{align}
\hbox{for}\ k>0:\qquad b&=\frac{-\i-(2N+1)\i\eE^k-\eE^{2k}\varphi-2\i\cos(k x_0)}{1+\eE^{2k}}+\frac{B-\overline
  B}{1+\eE^{2k}},\nonumber\\
\overline b&=\frac{\i+(2N+1)\i\eE^k+\eE^{2k}\varphi+2\i\cos(k x_0)}{1+\eE^{2k}}-\frac{B-\overline
  B}{1+\eE^{2k}}+C,\nonumber\\
\widetilde c&=B,\nonumber\\
\widetilde\Lambda&=\frac{-(2N+1)\i+\i\eE^k+\eE^{3k}\varphi+2\i\eE^k\cos(k
  x_0)}{\eE^k(1+\eE^{2k})}
+\frac{\eE^{2k}B+\overline  B}{1+\eE^{2k}}.\label{Beqkpos}
\end{align}
The last three equations of (\ref{eqkneg}) can be solved for $q_2$, $D$, $\widetilde\Lambda$ in terms of $B$, $\overline B$, $C$ and this inserted
into the first three equations gives
\begin{align}
\hbox{for}\ k<0:\qquad b&=\frac{-\i\eE^{2k}-(2N+1)\i\eE^k+\overline\varphi-2\i\eE^{2k}\cos(k x_0)}{1+\eE^{2k}}+\eE^{2k}\frac{B-\overline
  B}{1+\eE^{2k}}+C,\nonumber\\
\overline b&=\frac{\i\eE^{2k}+(2N+1)\i\eE^k-\overline\varphi+2\i\eE^{2k}\cos(k
  x_0)}{1+\eE^{2k}}-\eE^{2k}\frac{B-\overline B}{1+\eE^{2k}},\nonumber\\
\widetilde c&=\overline B,\nonumber\\
\widetilde\Lambda&=\frac{(2N+1)\i\eE^{3k}-\i\eE^{2k}+\overline\varphi-2\i\eE^{2k}\cos(k
  x_0)}{(1+\eE^{2k})}
+\frac{\eE^{2k}B+\overline  B}{1+\eE^{2k}}.\label{Beqkneg}
\end{align}
For many applications it is useful to consider these functions on shifted contours
\begin{equation}
a(x):=b(x-\i),\quad A(x):=B(x-\i),\quad
\overline a(x):=\overline b(x+\i),\quad \overline A(x):=\overline B(x+\i),
\end{equation}
because $a(x)$ and $\overline a(x)$ have a high order zero at $x=0$ rendering
the functions $A(x)$ and $\overline A(x)$ very small on large parts of the
real axis. Of course we must not attempt to calculate numerically the singular
$\log a(x)$ and $\log\overline a(x)$ directly. We will factor out of $a(x)$
and $\overline a(x)$ an explicit function yielding the zero of high order. It
will appear that in some of the formulas below, the shift $x\to x\pm\i$ should
be understood as a $x\to x\pm(1-\epsilon)\i$ with small positive $\epsilon$.

The Fourier transforms of $a$, $A$, $\overline a$, $\overline A$ and $b$, $B$,
$\overline b$, $\overline B$ are related by factors $\eE^{\pm k}$
\begin{equation}
a=\eE^k b,\quad A=\eE^k B,\quad \overline a=\eE^{-k}\overline b,\quad 
\overline A=\eE^{-k}\overline B.
\end{equation}
For these function we obtain the equations
\begin{align}
\hbox{for}\ k>0\qquad a&=\frac{-\i-(2N+1)\i\eE^k-\eE^{2k}\varphi-2\i\cos(k x_0)}{\eE^{k}+\eE^{-k}}+
\frac{\eE^{-k}}{\eE^{k}+\eE^{-k}}(A-\eE^{2k}\overline  A),\nonumber\\
\overline a&=\frac{\i\eE^{-2k}+(2N+1)\i\eE^{-k}+\varphi+2\i\eE^{-2k}\cos(k
  x_0)}{\eE^{k}+\eE^{-k}}
-\frac{\eE^{-k}}{\eE^{k}+\eE^{-k}}(\eE^{-2k}A-\overline  A)+\eE^{-k}C,\nonumber\\
\widetilde c&=\eE^{-k}A,\nonumber\\
\widetilde\Lambda&=\frac{-(2N+1)\i\eE^{-2k}+\i\eE^{-k}+\eE^{k}\varphi+2\i\eE^{-k}\cos(k
  x_0)}{\eE^k+\eE^{-k}}
+\frac{A+\overline  A}{\eE^{k}+\eE^{-k}},\label{Aeqkpos}
\end{align}
and
\begin{align}
\hbox{for}\ k<0\qquad a&=\frac{-\i\eE^{2k}-(2N+1)\i\eE^k+\overline\varphi-2\i\eE^{2k}\cos(k
  x_0)}{\eE^{k}+\eE^{-k}}
+\frac{\eE^{k}}{\eE^{k}+\eE^{-k}}(A-\eE^{2k}\overline A)+\eE^{k}C,\nonumber\\
\overline a&=\frac{\i+(2N+1)\i\eE^{-k}-\eE^{-2k}\overline\varphi+2\i\cos(k
  x_0)}{\eE^{k}+\eE^{-k}}
-\frac{\eE^{k}}{\eE^{k}+\eE^{-k}}(\eE^{-2k}A-\overline A),\nonumber\\
\widetilde c&=\eE^{k}\overline A,\nonumber\\
\widetilde\Lambda&=\frac{(2N+1)\i\eE^{2k}-\i\eE^{k}+\eE^{-k}\overline\varphi-2\i\eE^{k}\cos(k
  x_0)}{\eE^{k}+\eE^{-k}}
+\frac{A+\overline  A}{\eE^{k}+\eE^{-k}}.\label{Aeqkneg}
\end{align}
Finally we have to carry out the inverse Fourier transform which yields us
expressions for the derivatives of the functions $\log a(x)$,
$\log \overline a(x)$, $\log c(x)$ as sums of explicit functions and
convolution integrals of explicit functions with the derivatives
of the functions $\log A(x)$, $\log \overline A(x)$, $\log C(x)$. After
deriving these equations we take the integral and determine the constant of integration.

As a first step we carry out the inverse Fourier transforms of the explicit
functions. These functions are related to the digamma function $\psi$.
By use of the integral formula
\begin{equation}
\psi(z)=\int_0^\infty\left[\eE^{-t}-\frac 1{(1+t)^z}\right]\frac{dt}t,
\end{equation}
and the substitution $1+t=\eE^x$ we find
\begin{equation}
\psi(z+1/2)-\psi(z)=4\int_0^\infty dk\frac{\eE^{-(4z-1)k}}{\eE^k+\eE^{-k}}.
\end{equation}
This yields for a typical combination of terms appearing in the equation for $\log\Lambda(x)$
\begin{align}
\el(x,r):=&
\int_{-\infty}^0 dk \frac{\i\,\eE^{rk}}{\eE^k+\eE^{-k}}\eE^{\i kx}-
\int_0^\infty dk \frac{\i\,\eE^{-rk}}{\eE^k+\eE^{-k}}\eE^{\i kx}\nonumber\\
=&\frac \i{4}\left[\psi\left(\tfrac14(r+3+\i x)\right)+\psi\left(\tfrac14(r+1-\i x)\right)
-\psi\left(\tfrac14(r+1+\i x)\right)-\psi\left(\tfrac14(r+3-\i x)\right)\right].\label{defl}
\end{align}
Integrating this with respect to $x$ and suitably fixing the integration constant yields
\begin{align}
L(x,r)&:=\log\frac
{\Gamma\left(\tfrac 14(r+3+\i x)\right)\Gamma\left(\tfrac 14(r+3-\i x)\right)}
{\Gamma\left(\tfrac 14(r+1+\i x)\right)\Gamma\left(\tfrac 14(r+1-\i x)\right)}
+\log(4)= \log(x)+O\left(\frac 1{x^2}\right),
\end{align}
where the asymptotics is given for large arguments.

The next combination appears in the expressions for $\log a(x)$ and
$\log \overline a(x)$
\begin{align}
\kappa(x,r):=&
\int_{-\infty}^0 dk \frac{\eE^{rk}}{\eE^k+\eE^{-k}}\eE^{\i kx}+
\int_0^\infty dk \frac{\eE^{-rk}}{\eE^k+\eE^{-k}}\eE^{\i kx}\nonumber\\
  =&\frac 1{4}\left[\psi\left(\tfrac 14(r+3+\i x)\right)+\psi\left(\tfrac 14(r+3-\i x)\right)
-\psi\left(\tfrac 14(r+1+\i x)\right)-\psi\left(\tfrac 14(r+1-\i x)\right)\right].
\end{align}
The integral of this function with respect to $x$ is called $\alpha(x,r)$
\begin{align}
\alpha(x,r)&:=\i \log\frac
{\Gamma\left(\frac 14(r+3-\i x)\right)\Gamma\left(\frac 14(r+1+\i x)\right)}
{\Gamma\left(\frac 14(r+3+\i x)\right)\Gamma\left(\frac 14(r+1-\i x)\right)}
\quad\left(\to\pm{\frac\pi2} +{\cal O}\left(\frac 1{x}\right)\hbox{for}\ x\to\pm\infty\right).
\end{align}
Note that the introduced functions $\el(x,r)$, $L(x,r)$, $\kappa(x,r)$, $\alpha(x,r)$ are
real valued for real arguments $x$, $r$.

\begin{figure}[H]
\begin{center}
\includegraphics[width=0.8\textwidth]{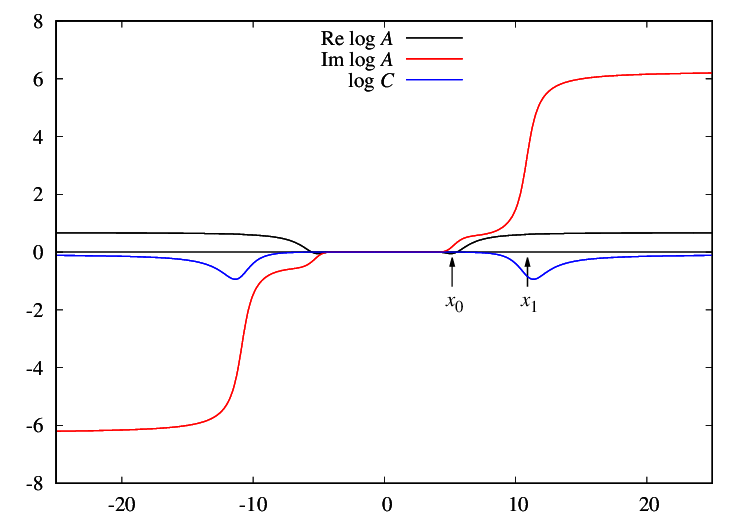}
\caption{Graphs of the functions $\log A$ and $\log C$ for the ground-state
  eigenvalue of the spin-$1/2$ XXX chain with boundary parameters $p=-0.6$,
  $q=-0.3$, $\xi=0.2$ and system size $N=10^3$.  The function
  $\log \overline A$ (not shown) is the complex conjugate of $\log A$. The
  function $\log C$ is real and even. The real (imaginary) part of $\log A$ is
  even (odd).  There are two important length ``scales'' in the system. First
  of all, there are the positions $\pm x_0$ of the zeros of the function
  $\log C$ (remember $C=1+c$ and $c(\pm x_0)=0$). Here
  $x_0=5.167..\simeq \tfrac2\pi\log N$. In the plot, the zeros can not be read
  off, because the function $\log C$ looks flat in $[-x_0,+x_0]$ where it
  actually takes amall positive values and noticeably negative values outside
  this interval.  In the same interval $[-x_0,+x_0]$ the real and imaginary
  parts of $\log A$ take vanishingly small values, outside they approach
  non-zero asymptotics. The behaviour of the imaginary part of $\log A$ is
  most interesting. After developing into some shoulder for $x>x_0$ it takes a
  steady increase for arguments $x$ in the vicinity of some value
  $x_1\sim 10.9\ (>x_0)$ and approaches $+2\pi$ for $x\to+\infty$. Note that
  in the vicinity of the same $\pm x_1$ the function $\log C$ shows minima.  }
\label{fig:winding}
\end{center}
\end{figure}

By use of these explicit functions we find for the eigenvalue function $\Lambda(x)$
\begin{align}
\log\Lambda(x)=&\log\widetilde\Lambda(x)+\log\lambda_0(x)\nonumber\\
=&(2N+1)L(x,2)-L(x,1)+L(x,p_1)+L(x,p_2)\nonumber\\
&+\log(x^2-x_0^2)-L(x-x_0,1)-L(x+x_0,1)\nonumber\\
&+e\ast(\log A+\log\overline A),\label{eigvalLambda3}
\end{align}
where the convolution $\ast$ of two functions $f, g$ is defined by
\begin{equation}
f\ast g(x)=\frac 1{2\pi}\int dy f(x-y)g(y),
\end{equation}
and the function $e(x)$ is given by the Fourier integral with explicit result
\begin{equation}
e(x):=\int_{-\infty}^\infty dk\frac{\eE^{\i
    kx}}{\eE^k+\eE^{-k}}=\frac{\frac{\pi}2}{\cosh\frac{\pi}2 x}.
\end{equation}
The integral expressions for the functions $\log a(x)$, $\log \overline a(x)$,
$\log c(x)$ are
\begin{align}
\log a(x)=& (2N+1)\log\tanh\left(\frac\pi 4x\right)+\frac\pi
            2\i+\log(2/\xi_1-1)-\i \alpha(x-\i,1)\nonumber\\
&+\i \alpha(x-\i,p_1)+\i \alpha(x-\i,p_2)-\i \alpha(x-x_0-\i,1)-\i \alpha(x+x_0-\i,1)\nonumber\\
&+K_{11}\ast\log A+K_{12}\ast\log \overline A+K_{13}\ast\log(C/C_\infty),\label{nliea}\\
\log \overline a(x)=& (2N+1)\log\tanh\left(\frac\pi 4x\right)-\frac\pi 2\i+\log(2/\xi_1-1)+\i \alpha(x+\i,1)\nonumber\\
&-\i \alpha(x+\i,p_1)-\i \alpha(x+\i,p_2)+\i \alpha(x-x_0+\i,1)+\i \alpha(x+x_0+\i,1)\nonumber\\
&+K_{21}\ast\log A+K_{22}\ast\log \overline A+K_{23}\ast\log(C/C_\infty),\label{nlieaa}\\
\log c(x)=& \log[x^2(x^2-x_0^2)]+K_{31}\ast\log A+K_{32}\ast\log \overline
A-4\log R,\label{nliec}
\end{align}
where the convolutions in (\ref{nliec}) are done with $\log A$ and
$\log \overline A$ evaluated on symmetric intervals $[-R,+R]$ with
$R\to\infty$. The kernel matrix is
\begin{equation}
K(x)=
  \left[ {\begin{array}{ccc}
  \kappa(x,1)  & -\kappa(x-(2-\epsilon)\i,1) & -\i/(x-\i)\\
  -\kappa(x+(2-\epsilon)\i,1) & \kappa(x,1) & \i/(x+\i)\\
   \i/(x+\i) & -\i/(x-\i) & 0\\
  \end{array} } \right],\qquad(\epsilon\ \hbox{small and positive}).
\end{equation}
Note that the careful prescription for the convolution integrals is necessary
because of the slow asymptotics of $K_{31}$ and $K_{32}$ and the curious
property that the functions $\log A$ and $\log \overline A$ show non-trivial
windings, see Fig.~\ref{fig:winding} for a system with size $N=10^3$, still to
be considered small on grounds that become clearer shortly. The equations
(\ref{nliea}), (\ref{nlieaa}), (\ref{nliec}) are non-linear integral equations
(NLIEs) for the functions $a$, $\overline a$, $c$, because $A=1+a$,
$\overline A=1+\overline a$, $C=1+c$. Numerical solutions are obtained by
iterative treatments and evaluations of the convolution integrals by the Fast
Fourier Transform. 
In parallel, the zero $x_0$ of $\Lambda(x)$ is found from solving $a(x_0+\i)=-1$ by evaluating the right hand side of (\ref{nliea}) off the real axis.

\section{\label{sec:level4} Bulk and surface energies, first numerical results}

There is a different and numerically better posed method to formulate the NLIEs
than done above. We introduce simple subtraction terms in the convolution part
such that the involved functions show vanishing asymptotics. And for the
compensation we use counter terms in the driving (source) terms of the
NLIEs. The proof of this form of NLIEs uses as intermediate step the
differentiated form of the NLIEs where no issues of convergence arise. There
the subtractions and compensation terms are introduced. They can be chosen to
be simply of rational function type, i.e.~the kernel $K$ convolved with
suitable rational functions yields rational functions.
After formulating the NLIEs for the derivatives $(\log a(x))'$,..., $(\log
A(x))'$,... the NLIEs are integrated with respect to the argument $x$ and the
integration constants are fixed by considering the limit $x\to\infty$.
The result is
\begin{equation}
\left(\begin{matrix}
\log a\cr \log \overline a\cr \log c
\end{matrix}\right)=d+K\ast \left(\begin{matrix}
\log (A/A_\infty)-\log\left({\displaystyle \frac{x-x_{r+}}{x-x_{r-}}\cdot\frac{x-x_{l+}}{x-x_{l-}}}\right)\cr
\log (\overline A/\overline A_\infty) -\log\left({\displaystyle \frac{x-x_{r-}}{x-x_{r+}}\cdot\frac{x-x_{l-}}{x-x_{l+}}}\right)\cr
\log(C/C_\infty)
\end{matrix}\right),\label{NLIEcounterterms}
\end{equation}
where the asymptotic values $A_\infty=1+a_\infty$, $\overline A_\infty=1+\overline a_\infty$ and $C_\infty=1+c_\infty$ are easily obtained from (\ref{limits}) and $a_\infty=b_\infty$ etc.
The parameters
$x_{r+}$ and $x_{r-}$ are complex numbers with positive real part and
positive resp.~negative imaginary parts. And $x_{l+}$ and $x_{l-}$ are defined similarly
with negative real parts. The introduced functions subtract the winding
behaviour observed in the functions $\log A$ and $\log \overline A$. Now the
inhomogeneity $d$ is a tuple of three functions containing the counter terms
\begin{align}
d_1(x)&=(2N+1)\log\tanh\left(\frac\pi 4x\right)+\frac\pi 2\i
            -\i \alpha(x-\i,1)\nonumber\\
&+\i \alpha(x-\i,p_1)+\i \alpha(x-\i,p_2)-\i \alpha(x-x_0-\i,1)-\i \alpha(x+x_0-\i,1)\nonumber\\
&+\log\left({\displaystyle a_\infty\frac{x-x_{r+}-2\i}{x-x_{r-}}\cdot\frac{x-x_{l+}-2\i}{x-x_{l-}}}\right),\\
d_2(x)&=(2N+1)\log\tanh\left(\frac\pi 4x\right)-\frac\pi 2\i
         +\i \alpha(x+\i,1)\nonumber\\
&-\i \alpha(x+\i,p_1)-\i \alpha(x+\i,p_2)+\i \alpha(x-x_0+\i,1)+\i \alpha(x+x_0+\i,1)\nonumber\\
  &+\log\left({\displaystyle \overline a_\infty\frac{x-x_{r-}+2\i}{x-x_{r+}}\cdot\frac{x-x_{l-}+2\i}{x-x_{l+}}}\right),\\
d_3(x)&=  \log\left({\displaystyle c_\infty\frac{x^2(x^2-x_0^2)}{(x-x_{r-}+\i)(x-x_{r+}-\i)(x-x_{l-}+\i)(x-x_{l+}-\i)}}\right).
\end{align}
The concrete values of $x_{r\pm}$ and $x_{l\pm}$ drop out of the calculations
and affect at best the accuracy of the calculations.
For practical purposes we choose for these numbers
\begin{equation}
  x_{r\pm}=\tilde x_1\pm \i \,\delta \qquad   x_{l\pm}=-\tilde x_1\pm \i \,\delta,
\end{equation}
with some $\delta>0$ and $\tilde x_1$ is an estimate of the location of the transition of the
imaginary part of $\log A$ from small resp.~practically zero values to $+2\pi$ as explained in
Fig.~\ref{fig:winding}.

A natural strategy for numerical iterations of the NLIEs is to use functions as
initial data that have qualitative behaviour in common with solutions obtained
from solving the Bethe ansatz equations directly. This of course is possible
only for small system sizes ($N\sim 10$). For such systems the
  function $a(x)$ takes vanishingly small values for arguments $x$ close to
  0. Increasing $x$ beyond values of about $\tfrac2\pi\log N$ leads to a
  noticeable increase of the absolute value of $a(x)$ under some small angle
  to the real axis. Then for larger values of the argument, at some $x_1$, the
  values of $a(x)$ move sharply into the complex plane around the point $-1$
  which is encircled exactly once in counter-clockwise manner. This is the
  reason for the sharp increase of Im$(\log A)$ in Fig.~\ref{fig:winding}.

By use of qualitatively similar initial data for the iterative treatment of
the NLIEs we found convergence for much larger system sizes like that shown in
Fig.~\ref{fig:winding} for $N=10^3$. However, increasing the system size
further will ultimately and independently of the chosen boundary parameters
lead to the loss of convergence. 

Independent of the system size we found that the zero $x_0$ of the function
$c(x)$ scales like $\tfrac2\pi\log N$. The point $x_0$ also separates ranges
of the argument $x$ for which $a(x)$ takes vanishingly small values, due to
the leading $\tanh^{2N+1}(\tfrac\pi4x)$ factor resulting from the NLIEs, and
larger values of $x$ with $a(x)$ being of order 1. From the NLIEs it is also
obvious and actually necessary that $a(x)$ describes the encircling of $-1$ and
hence $\log A$ shows two times an increase of the imaginary part by
$2\pi\i$. Then the function $\log c(x)$ has an asymptotic behaviour that is no
longer of order $\log x^4$, but instead it is now the logarithm of a rational
function with order 4 polynomials in the numerator and in the denominator. It
is or seems absolutely natural that the encircling of $-1$ by $a(x)$ happens at arguments
close to a value $x_1$ which is larger than $x_0$: only for $x>x_0$ the
values $a(x)$ move away from 0.

For large system sizes we simply did not find any suitable initial data set
with the winding occurring at some $x_1>x_0$ whether we chose $x_1$ much
larger than $x_0$ or of similar size. For a failure of convergence many
scenarios are conceivable, but the resolution of the problem appeared in the
least expected manner. The winding happens in the ``forbidden region''
$[-x_0,+x_0]$ where the values of the function $a(x)$ seem to be tied to 0.

For small system sizes there are Bethe roots, zeros of
  $q_1(x)$ and $q_2(x)$, with largest positive (negative) real parts that form
  a complex conjugate pair of a pole and a zero of the function
  $A(x)$. Following the function $A(x)$ for real values of $x$ straight
  through the complex conjugate pole/zero pair results into $A(x)$ performing
  a loop around 0. The scaling of this extremal pole/zero pair with increasing
  size $N$ is difficult to follow. Increasing $N$ from very small values shows
  a motion of the pairs away from the origin.

\begin{figure}
\begin{center}
  \includegraphics[width=0.8\textwidth]{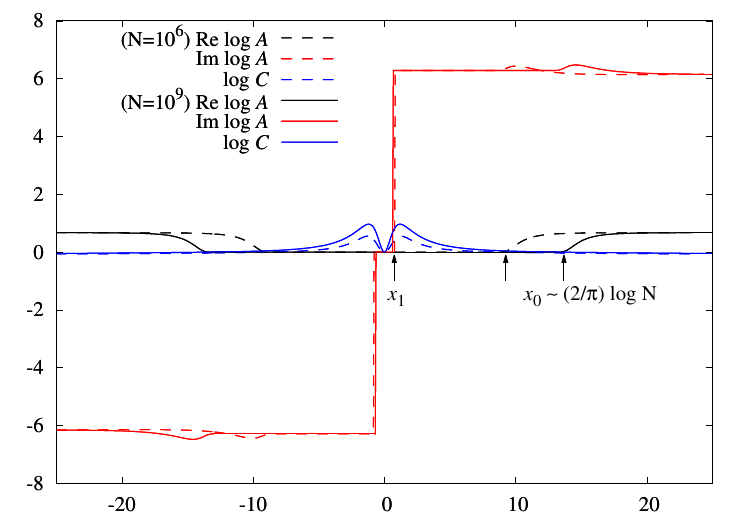}
  \caption{Plot of the functions $\log A$ and $\log C$ for the ground-state
    eigenvalue of the spin-$1/2$ XXX chain with parameters and colour coding
    as in Fig.~\ref{fig:winding} for larger system sizes $N=10^6$ (dashed
    lines) $N=10^9$ (solid lines).
    The function $\log C$ is zero at $\pm x_0$ with value
    $x_0=9.24...(13.68...) \simeq \tfrac2\pi\log N$ for $N=10^6\ (10^9)$. We
    note the qualitative differences to Fig.~\ref{fig:winding}. Here, the point
    $x_1$, at which the imaginary part of $\log A$ increases sharply (!) from $0$
    to $+2\pi$, has a value less (!) than $x_0$. Within the resolution of the
    figure this is a step function. The point $x_1$ is rather well defined
    with value $x_1=0.83... (0.64...)$ for $N=10^6\ (10^9)$. Note that $x_0$
    increases and $x_1$ decreases with $N$. The function $\log C$ takes
    positive and noticeably large values within the interval $[-x_0,+x_0]$ and
    rather flat negative values outside. In the neighbourhood of $\pm x_1$ of
    the sharp transitions of the imaginary part of $\log A$ the function
    $\log C$ shows maxima.}
\label{fig:windingLlarge}
\end{center}
\end{figure}

We postulate that for some system size this trend reverses and the extremal
pairs move back to the origin from which they keep a finite distance in the limit
$N\to\infty$. The condition for this to happen is that the separation of the
zero and pole in the pair is approaching 0 exponentially fast. Under this
condition, the above decribed winding happens so fast that the convolution
integrals produce contributions that cancel the leading
$(2N+1)\log\tanh\tfrac\pi4x$. On the basis of this reasoning we found
solutions of the NLIEs with much larger system sizes like those illustrated in 
Fig.~\ref{fig:windingLlarge}.

Next we turn to the calculation of finite size-corrections.
From the eigenvalue expression (\ref{eigvalLambda3}) the
energy is obtained by
\begin{equation}
\label{EE}
\begin{aligned}
E=-2\i\frac\partial{\partial x}\log\Lambda(x)\Big|_{-\i}-N
=&-(2N+1)2\i\, \el(-\i,2)-N+2\i\,[\el(-\i,1)-\el(-\i,p_1)-\el(-\i,p_2)]\\
&+2\i\left[\frac1{\i+x_0}+\frac1{\i-x_0}+\el(-\i-x_0,1)+\el(-\i+x_0,1)\right]\\ 
&-2\i\left[e'\ast(\log A+\log\overline A)\big|_{-\i}\right]\,.
\end{aligned}
\end{equation}
By use of (\ref{defl}) we obtain
\begin{equation}
2\i\,\el(-\i,r)=\psi\left(\frac{r+2}4\right)-\psi\left(\frac{r}4\right)-\frac2r.\label{lspec}
\end{equation}
Here and in the following calculations we make use of the functional equations and
special argument identities
\begin{align*}
\psi(x+1)&=\psi(x)+\frac1x,\\
\psi(1-x)-\psi(x)&=\pi\cot(\pi x), \quad\psi(\tfrac34)-\psi(\tfrac14)=\pi,\\
\psi(1)-\psi(\tfrac12)&=2\log2.
\end{align*}

With these identities the first line of (\ref{EE}) gives the known bulk and boundary contributions to the ground state energy of the antiferromagnetic XXX spin-$1/2$ chain with boundary fields in the thermodynamic limit \cite{Hult39,FrZv97b}:
\begin{equation}
\begin{aligned}
 E_0 =& N-1 + \frac2{p_1} + \frac2{p_2} +N\epsilon_\infty + f_\infty +\mathcal{O}\left(\frac1N\right)\,,\\
    &\quad \epsilon_\infty = -4\log2\,,\quad
      f_\infty = \pi-2\log2
 + \sum_{k=1,2} \left(\psi\left(\frac{p_k}4\right)-\psi\left(\frac{p_k+2}4\right)\right)\,.
\end{aligned}
\end{equation}
%
Similarly, one finds
\begin{align}
  &2\i\left[\frac1{\i+x_0}+\frac1{\i-x_0}+\el(-\i-x_0,1)+\el(-\i+x_0,1)\right]
\nonumber\\
  &\qquad=\pi\cot\tfrac\pi4(1+\i x_0)+\pi\cot\tfrac\pi4(1-\i x_0)=\frac{2\pi}{\cosh\frac\pi2x_0}
  =4 e(x_0)\,.
\end{align}
In summary the ground state energy of the spin chain with non-diagonal boundary fields is 
\begin{equation}
  E= E_0 
+4 e(x_0)-2\i\,\left[e'\ast(\log A+\log\overline A)\big|_{-\i}\right]\,.\label{Ef}
\end{equation}
Note that $E-E_0=\mathcal{O}(1/N)$, because $e(x_0)\simeq
\tfrac\pi N$ and the function $\log A(x)+\log\overline A(x)$ takes
non-negligeable values only for arguments $|x|>x_0$.
The calculation of the $1/N$ corrections involves the dilog-trick and quantitative
calculations for the functions $A$, $\overline A$, $C$ in the scaling limit
$N\to\infty$. These results will be communicated in separate publications.

\section{\label{sec:level5}Conclusion}
The XXX spin chain with non-diagonal boundary fields stands as a prominent yet
challenging example of quantum integrable systems with non-trivial boundaries
\cite{DeVega1994,Nepo02,Nepo03,fra11,Cao2013,Wang2016, Nepomechie2013,Belliard2015,Belliard2015b,Crampe2017, Avan2015ModifiedXXZ,FrSW08,Niccoli2012NonDiagonalXXZ, Faldella2014CompleteSpectrum,Dong2023}. Despite
its relevance for the theory of integrable systems the exact solution has
remained a longstanding problem. A great part of a satisfactory solution is
realized by the derivation of inhomogeneous $T-Q$ relations and the corresponding Bethe equations, discovered by the
off-diagonal Bethe Ansatz (ODBA) \cite{Cao2013,Wang2016} and further
understood and extended in other work
\cite{Nepomechie2013,Belliard2015,Belliard2015b,Crampe2017,Avan2015ModifiedXXZ,FrSW08,Niccoli2012NonDiagonalXXZ,Faldella2014CompleteSpectrum}. The analytic treatment of these equations has so far been limited to the thermodynamic limit \cite{Li2014}, without retaining terms that characterize finite-size corrections. While bulk and surface energies, as well as certain excitations, were analyzed in \cite{Dong2023,Li2014}, finite-size effects remained elusive. In this work, we successfully construct nonlinear integral equations (NLIEs) to tackle this challenge.

A key advancement in our work is the introduction of the function \(c(x)\),
which accounts for the inhomogeneous term in the \(T\)-\(Q\) relation. This
complements the classical functions \(a(x)\) and \(\bar{a}(x)\) and allows for
the complete description of the system. When the boundary fields are taken
parallel the NLIEs simplify significantly, reducing to two coupled equations
without long-range kernel terms. These simplified equations are
computationally efficient to solve. However, for non-parallel boundary fields,
i.e.~non-zero values of the parameter \(\xi\) the situation is different. First of all, we
observe the winding phenomenon, the functions \(\log A(x)\) and
\(\log\bar A(x)\) show sudden changes by $2\pi\i$ at some characteristic
scale $x_1$ of the argument. At a rather different value $x_0$ the function
\(\log C(x)\) turns zero. Second, we realized that the two scales $x_0$ and $x_1$ are
independent. We succeeded in obtaining explicit numerical results for large
and small values of the ratio $x_1/x_0$ which are taken for small and large
system size $N$. 

This study represents the first step in our broader project to explore the
conformal properties of the Heisenberg spin chain through the lens of
NLIEs. The large $N$ results allow for future analytical study of the finite
size properties of the system with non-parallel boundary fields.  The next
step will be the derivation of a suitable scaling limit of the
NLIEs. Preliminary studies have yielded a simplified kernel matrix still with
the same long-range terms, but all regular terms being simplified to
delta-functions. The full understanding of this and the combined
analytical-numerical investigations require additional work that will be
published elsewhere.

By starting with the isotropic XXX model, we have developed a framework
that paves the way for generalizations to the XXZ spin chain with arbitrary
open boundary conditions. Our approach is designed to address the challenges
posed by the loss of \(U(1)\) symmetry in these systems and to provide a
robust method for analyzing finite-size corrections 
and
related properties.

Ultimately, we aim to contribute not only to the theoretical understanding of
integrable models but also to their application in broader contexts, such as
statistical mechanics and condensed matter physics, where boundary effects
play a critical role.

\section*{Acknowledgment} 
X.Z. acknowledges financial support from the National Natural Science
Foundation of China (No. 12204519) and from the Alexander von Humboldt
Foundation. A.K., D.W. and H.F. acknowledge financial support by Deutsche
Forschungsgemeinschaft through FOR 2316. 
A.K. acknowledges hospitality by the
Innovation Academy for Precision Measurement Science and Technology, Wuhan, and funding by the Chinese Academy of
Sciences President's International Fellowship Initiative, Grant No. 2024PVA0036. 


\printbibliography

\end{document}